\documentstyle[tighten,preprint,prd,aps]{revtex}

\begin{document}

\draft

\title{
The Nordtvedt Effect in Rotational Motion
}

\author{Sergei A. Klioner and Michael Soffel}
\address{
           Lohrmann Observatory,
           Dresden Technical University,\\
           Mommsenstra\ss e 13,
           D-01062 Dresden, Germany
}

\date{\today}

\maketitle

\begin{abstract}
Rotational motion of extended celestial bodies is discussed in the
framework of the Parametrized Post-Newtonian (PPN) formalism with the
two parameters $\gamma$ and $\beta$. A local PPN reference system of a
massive extended body being a member of a system of $N$ massive
extended bodies is constructed.  In the local PPN reference system the
external gravitational field manifests itself only in the form of tidal
potentials. Rotational equations of motion, which are then derived in
the local reference system, reveal a special term in the torque
analogous to the Nordtvedt effect in the translational equations of
motion: it is proportional to $4\beta-\gamma-3$, to the acceleration of
the body relative to the global PPN reference system and to some
quantity characterizing the distribution of inertial gravitational
energy within the body. This term is a direct consequence of the
violation of the Strong Equivalence Principle in alternative theories
of gravitation.
\end{abstract}

\section{Introduction}

In present paper we deal with rotational motion of a
celestial body being a member of a system of $N$ extended bodies in the
PPN formalism. It is well known that rotational motion of non-isolated
extended celestial bodies is a subtle point.  In General Relativity
this problem has been treated in the post-Newtonian approximation by
Fock \cite{fock} who used one global reference system to describe both
translational motion of $N$ extended rotating bodies and their
rotational motion. The resulting equations of motion derived by Fock
showed complicated couplings between translational motion of the bodies
and their rotation. However, most of these couplings were only spurious
coordinate effects which were later shown to vanish when using
physically adequate local reference systems to describe the local
dynamics (e.g., rotation) of each body of the system (see \cite{dsx:93}
for further comments and references).  The idea of the local reference
systems is to introduce a reference system for a body in which the
influence of external masses is effaced as much as possible.  In
General Relativity the local reference systems have been discussed in
great details by Brumberg and Kopeikin
\cite{brum:kopej:89b,kopej:88} (see also \cite{klio:voi:93})
and Damour, Soffel and Xu \cite{dsx:91,dsx:92,dsx:93}. In particular it
has been shown that the local reference systems in General Relativity
have two properties:

\begin{itemize}
\item[({\bf A})] the gravitational field of external bodies is
represented in the form of tidal potentials being ${\cal O}({\bf X}^2)$,
where $X^i$ are local spatial coordinates;

\item[({\bf B})] the internal gravitational field of the body coincides
formally with the gravitational field of a corresponding isolated source
provided that the tidal influence of the external matter is neglected.
\end{itemize}

The first attempt to use a version of such a local reference system to
study the rotational motion of an extended body has been undertaken by Voinov
\cite{voi:88}. Major progress to solve the problem in the Einsteinian
post-Newtonian theory was achieved by Damour, Soffel and Xu
\cite{dsx:93} who used their DSX formalism aimed at constructing the
local reference systems and derived the rotational equations of motion of
each body of an $N$-body system with full multipole structure.

The aim of this paper is to derive rotational equations of motion of an
arbitrarily-shaped extended body being a member of a system of $N$
arbitrarily-shaped extended bodies in the framework of the PPN
formalism with the two Eddington parameters $\beta$ and $\gamma$.  The
PPN formalism is a phenomenological scheme giving in generic
parametrized form the post-Newtonian metric tensor for an isolated
material source in a wide class of metric theories of gravitation.
Many aspects of testing General Relativity in the weak-field
slow-motion regime are based upon the PPN formalism. On the other hand,
the theory of local reference systems is also proved to be very
important for physically meaningful modeling of observational data.
Therefore, it is quite important to elaborate a theory of local
reference systems in the framework of the PPN formalism
(see also \cite{klio:sof:97}).

In full detail our theory of local PPN reference systems will be
discussed elsewhere. Here, we confine ourselves to a brief description
of the theory and to one of its results: rotational equations of motion
of a body relative to its local reference system. These equations
reveal an interesting physical effect. They contain an additional
torque analogous to the Nordtvedt effect in the translational equations
of motion: the torque is proportional to the Nordtvedt parameter
$\eta=4\beta-\gamma-3$, to the acceleration of the body relative to the
global PPN reference system and finally to some integral over the
volume of the body characterizing the distribution of the internal
gravitational energy inside the body.

\section{Global PPN reference system}

We begin with the global PPN metric tensor of a isolated
$N$ body system in the form (see, e.g., \cite{will:93})

\begin{eqnarray}\label{metric-BRS}
g_{00}&=&-1+{2\over c^2}\,w(t,{\bf x})-{2\over c^4}
\,\beta\, w^2(t,{\bf x})+{\cal O}(c^{-5}),
\nonumber \\
g_{0i}&=&-{2(1+\gamma)\over c^3}\,w^i(t,{\bf x})+{\cal O}(c^{-5}),
\nonumber \\
g_{ij}&=&\delta_{ij}\left(1+{2\over c^2}\,\gamma\,
w(t,{\bf x})\right)+{\cal O}(c^{-4}),
\end{eqnarray}
\noindent
where
\begin{eqnarray}
\label{w}
w&=&G\,\int\,\sigma(t,{\bf x}')\,{1\over |{\bf x}-{\bf x}'|}\,d^3x'+
    {1\over 2c^2}\,G\,{\partial^2\over\partial t^2}\,
\int \sigma(t,{\bf x}')\,|{\bf x}-{\bf x}'|\,d^3x'
+{\cal O}(c^{-4}),
\\ \label{wi}
w^i&=&G\,\int\,\sigma^i(t,{\bf x}')\,{1\over |{\bf x}-{\bf x}'|}\,d^3x'
+{\cal O}(c^{-2}),
\\ \label{sigmas}
\sigma&=&
{1\over c^2}\left(T^{00}+\gamma T^{kk}+
{1\over c^2} T^{00}\,(3\gamma-2\beta-1)\,w\right)+{\cal O}(c^{-4}),
\qquad
\sigma^i={1\over c} T^{0i}+{\cal O}(c^{-2})
\end{eqnarray}
\noindent
and $T^{\alpha\beta}$ is the energy-momentum tensor in the global PPN
reference system $(t,x^i)$.  We retain here only two PPN parameters:
$\gamma$ and $\beta$. This simplest version of the PPN formalism
covers, nevertheless, most viable theories of gravitation.
For some technical reasons we prefer to work not in the standard PPN
gauge, but in a different one, the difference affecting only the
coordinate time $t$
\begin{equation}\label{gauges}
t_{\rm PPN}=t-{1\over c^4}\,{\partial\over\partial t}\,\chi
+{\cal O}(c^{-5}),
\end{equation}
\noindent
where $t_{\rm PPN}$ is coordinate time of the global PPN reference
system in the standard PPN gauge and
\begin{equation} \label{chi}
\chi={1\over 2}\,G\,\int \sigma(t,{\bf x}')\,
|{\bf x}-{\bf x}'|\,d^3x'.
\end{equation}
\noindent
For $\gamma=\beta=1$ the reference system $(t,x^i)$ is harmonic.

We have written the metric tensor of the global PPN reference system
without specifying an explicit form of the energy-momentum tensor.
This is no more than a formal way to specify the global PPN metric
tensor.  The metric (\ref{metric-BRS})--(\ref{sigmas}) coincides with
the version of the PPN formalism described in \cite{will:93} as well as
with the version discussed in \cite{dsx:73}, provided that only two
parameters $\gamma$ and $\beta$ are retained in both versions.
Effectively we consider only those theories of gravitation which
produce the metric (\ref{metric-BRS})--(\ref{sigmas}) in the first
post-Newtonian approximation, the energy-momentum tensor being left
unspecified.

We consider the material system to consist of $N$ bodies which represent
simply spatially bounded blobs of matter.  Potentials $w$ and $w^i$ are
defined by (\ref{w})--(\ref{wi}) as volume integrals over the whole
space. Using this fact, we split the area of integration into the
volume $V$ of the body, for which we want to construct a local
reference system, and the remaining part of space. Thus, we split $w$ and
$w^i$ into internal potentials (potentials of the body under
consideration) and external ones (potentials due to the other bodies):
\begin{eqnarray}\label{BRS:split}
w(t,{\bf x})&=&w_{\rm E}(t,{\bf x})+{\overline{w}}(t,{\bf x}),
\nonumber \\
w^i(t,{\bf x})&=&w^i_{\rm E}(t,{\bf x})+{\overline{w}}^i(t,{\bf x}).
\end{eqnarray}

\section{Local PPN reference system}

From a physical point of view any reference system covering a
space-time region under consideration can be used to describe physical
phenomena within that region. Some reference systems, however, offer
a simpler mathematical description for physical laws and make the
character of physical laws more obvious.  It is well known that, for
example, for a massless observer one can construct a local reference
system where gravitational fields appear only in tidal form.  Due to
the works of Brumberg and Kopeikin
\cite{brum:kopej:89b,kopej:88} and Damour, Soffel and Xu
\cite{dsx:91,dsx:92,dsx:93} we know that in the first post-Newtonian
approximation of General Relativity it is
possible to construct an analogue of such a local reference system for a
massive extended body and that those local reference systems have two
properties {\bf A} and {\bf B} listed in the Introduction. It is clear that
the possibility to satisfy in General Relativity both properties
simultaneously is closely related with the Equivalence Principle. It
can be expected apriori and will be seen below that in alternative
theories of gravitation where the Equivalence Principle is violated
properties {\bf A} and {\bf B} cannot be satisfied simultaneously.

We assume the metric tensor of the local PPN reference system
$(T,X^a)$ of a selected body to be of the form
\begin{eqnarray}\label{metric:GRS}
G_{00}&=&-1+{2\over c^2}\,W(T,{\bf X})-{2\over c^4}
\,\beta\,W^2(T,{\bf X})+{\cal O}(c^{-5}),
\nonumber \\
G_{0a}&=&-{2(1+\gamma)\over c^3}\,W^a(T,{\bf X})+{\cal O}(c^{-5}),
\nonumber \\
G_{ab}&=&\delta_{ab}\left(1+{2\over c^2}\,\gamma\,
W(T,{\bf X})\right)+{\cal O}(c^{-4}),
\end{eqnarray}
\noindent
and the local gravitational potentials $W$ and $W^a$ to admit a
representation in the form
\begin{eqnarray}\label{GRS:split}\label{W-split}
W(T,{\bf X})&=&W_{\rm E}(T,{\bf X})+Q_a(T)\,X^a+W_{\rm T}(T,{\bf X})
+{1\over c^2}\,\Psi(T,{\bf X}),
\\ \label{Wi-split}
W^a(T,{\bf X})&=&W^a_{\rm E}(T,{\bf X})
+{1\over 2}\,\varepsilon_{abc}\,C_b(T)\,X^c
+W^a_{\rm T}(T,{\bf X}).
\end{eqnarray}
\noindent
Here, the local internal gravitational potentials $W_{\rm E}$ and $W_{\rm E}^a$ have
the same functional form as their global counterparts $w_{\rm E}$ and
$w_{\rm E}^i$, but all quantities should be taken in the local reference
system, i.e.
\begin{eqnarray}
\label{W}
W_{\rm E}&=&G\,\int_V\,\Sigma(T,{\bf X}')\,{1\over |{\bf X}-{\bf X}'|}\,d^3X'
    +{1\over 2c^2}\,G\,{\partial^2\over\partial T^2}\,
\int_V\,\Sigma(T,{\bf X}')\,|{\bf X}-{\bf X}'|\,d^3X'+{\cal O}(c^{-4}),
\\ \label{Wi}
W^a_{\rm E}&=&G\,\int_V\,\Sigma^a(T,{\bf X}')\,
{1\over |{\bf X}-{\bf X}'|}\,d^3X'
+{\cal O}(c^{-2}),
\\ \label{Sigmas-GRS}
\Sigma&=&
{1\over c^2}\left({\cal T}^{00}+\gamma {\cal T}^{aa}+
{1\over c^2} {\cal T}^{00}\,(3\gamma-2\beta-1)\,W\right)+{\cal O}(c^{-4}),
\qquad
\Sigma^a={1\over c} {\cal T}^{0a}+{\cal O}(c^{-2}),
\end{eqnarray}
\noindent
where ${\cal T}^{\alpha\beta}$ is the energy-momentum tensor in the
local reference system.  The external potentials $W_{\rm T}$ and
$W^a_{\rm T}$ represent tidal field of the other bodies of the system
and are assumed to be $\sim {\cal O}({\bf X}^2)$. It is also assumed
that $(W_{\rm T},W^a_{\rm T})$ are functions of
$(\overline{w},\overline{w}^i)$ and their derivatives as well as of the
trajectory of the origin of the local reference system relative to the
global one. Two arbitrary functions $Q_a(T)$ and $C_a(T)$ have a clear
physical meaning which will be discussed below.  Finally, the function
$\Psi(T,{\bf X})$ is some unknown function containing internal
potentials of the central body which appear in the local PPN reference
system in addition to $W_{\rm E}$ and $W^a_{\rm E}$. Clearly the
appearance of $\Psi$ is related with a violation of the Equivalence
Principle which make it impossible to satisfy simultaneously properties
{\bf A} and {\bf B} formulated in the Introduction. By assuming that
$W_{\rm T}$ and $W^a_{\rm T}$ are $\sim {\cal O}({\bf X}^2)$ we assume
that property {\bf A} is satisfied. Therefore, property {\bf B} is
violated which results in the appearance of $\Psi$.

Then, the results of the Brumberg-Kopeikin and DSX formalisms (see,
e.g., Theorems 1 and 2 of \cite{dsx:91}) allow us to
write the transformations between the global and local reference
systems in the form
\begin{eqnarray}
\label{trans:time}
&&T=t-{1\over c^2}\left(A+v_{\rm E}^ir_{\rm E}^i\right)+
{1\over c^4}\left(B+B^i r_{\rm E}^i+ B^{ij} r_{\rm E}^i r_{\rm E}^j
+C(t,{\bf x})
\right)+{\cal O}(c^{-5}),
\\ \label{trans:space}
&&X^a=R^a_{\,j}\left(r_{\rm E}^j+{1\over c^2}
\left({1\over 2}v_{\rm E}^j\,v_{\rm E}^k r_{\rm E}^k+D^{jk}\,r_{\rm E}^k
+\gamma\,\left(
r_{\rm E}^j a_{\rm E}^k r_{\rm E}^k-{1\over 2}\,a_{\rm E}^j
r_{\rm E}^2\right)\right)\right)
+{\cal O}(c^{-4}),
\end{eqnarray}
where $r_{\rm E}^i=x^i-x_{\rm E}^i(t)$, $x_{\rm E}^i(t)$ is the
coordinates of the origin of the local reference system relative to the
global one, and $v_{\rm E}^i=dx_{\rm E}^i/dt$ and $a_{\rm
E}^i=d^2x_{\rm E}^i/dt^2$ are its velocity and acceleration,
respectively. The functions $A(t)$, $B(t)$, $B^i(t)$, $B^{ij}(t)=
B^{ji}(t)$, $D^{ij}(t)=D^{ji}(t)$, $R^a_{\ j}(t)$ (being orthogonal
matrix) and $C(t,{\bf x})\sim{\cal O}(r_{\rm E}^3)$ are some unknown
functions.

The transformation rule
\begin{equation}\label{matching}
g_{\alpha\lambda}(t,{\bf x})=
{\partial X^\mu\over \partial x^\alpha}\,
{\partial X^\nu\over \partial x^\lambda}\,
G_{\mu\nu}(T,{\bf X})
\end{equation}
enables us then to derive or constrain the unknown functions from
both the local PPN metric tensor (\ref{metric:GRS})--(\ref{Sigmas-GRS})
and the transformations (\ref{trans:time})--(\ref{trans:space})
\begin{eqnarray}
\label{A}
&&{\dot A}(t)={1\over 2}v_{\rm E}^2+{\overline{w}}({\bf x}_{\rm E}),
\\
\label{B}
&&{\dot B}(t)=-{1\over 8}v_{\rm E}^4
+2(\gamma+1)\,v_{\rm E}^i\,{\overline{w}}^i({\bf x}_{\rm E})
-\left(\gamma+{1\over 2}\right)\,v_{\rm E}^2\,{\overline{w}}({\bf x}_{\rm E})
+\left(\beta-{1\over 2}\right)\,
{\overline{w}}^2({\bf x}_{\rm E}),
\\
\label{B^i}
&&B^i(t)=-{1\over 2}v_{\rm E}^2\,v^i_{\rm E}
+2(1+\gamma)\,{\overline{w}}^i({\bf x}_{\rm E})
-(2\gamma+1)\,v_{\rm E}^i\,{\overline{w}}({\bf x}_{\rm E}),
\\
\label{D^ij}
&&D^{ij}=\gamma\,\delta^{ij}\,\overline{w}({\bf x}_{\rm E}),
\\
\label{B^ij}
&&B^{ij}(t)=-v_{\rm E}^{(i} R^a_{\ j)} Q_{a}+
(1+\gamma)\,{\overline{w}}^{(i,j)}({\bf x}_{\rm E})
-\gamma\,v_{\rm E}^{(i}\,{\overline{w}}^{,j)}({\bf x}_{\rm E})
+{1\over 2}\gamma\,\delta^{ij}\,{\dot{\overline{w}}}({\bf x}_{\rm E}),
\\
\label{Rij-Ca}
&&c^2\,R^a_{\ i}\,\dot R^a_{\ j}=-(1+\gamma)\,\varepsilon_{ijk} R^a_{\ k}C_a
\nonumber\\
&&\phantom{c^2\,R^a_{\ i}\,\dot R^a_{\ j}=}
-2(1+\gamma)\,\overline{w}^{[i,j]}({\bf x}_{\rm E})
+(1+2\gamma)\,v_{\rm E}^{[i}\,\overline{w}_{,j]}({\bf x}_{\rm E})
+v_{\rm E}^{[i}R^a_{\ j]}Q_a+{\cal O}(c^{-2}),
\\
\label{C:laplace}
&&C_{,ii}=(\gamma-2)\,{\dot{a}_{\rm E}}^k\,r_{\rm E}^k,
\\
\label{aei}
&&a_{\rm E}^i(t^*)={\overline{w}}_{,i}(t^*,{\bf x}_{\rm E}(t^*))
-R^a_{\ j}Q_a(T)
\left(\delta^{ij}-{1\over c^2}\left(
v_{\rm E}^2\,\delta^{ij}
+(2+\gamma)\,\overline{w}({\bf x}_{\rm E})\,\delta^{ij}
+{1\over 2} v_{\rm E}^i v_{\rm E}^j
\right)\right)
\nonumber \\
&&
\phantom{a_{\rm E}^i(t^*)=}
+{1\over c^2}\biggl(
2(1+\gamma)\,\dot{{\overline{w}}}^i({\bf x}_{\rm E})
+\left(\gamma v_{\rm E}^2
-2(\gamma+\beta)\,{\overline{w}}({\bf x}_{\rm E})\right)
\overline{w}_{,i}({\bf x}_{\rm E})
-(2\gamma+1)\,v_{\rm E}^i\,\dot{{\overline{w}}}({\bf x}_{\rm E})
\nonumber \\
&&
\phantom{a_{\rm E}^i(t^*)=+{1\over c^2}\biggl(}
-2(1+\gamma)\,v_{\rm E}^j\,{\overline{w}}^j_{,i}({\bf x}_{\rm E})
-v_{\rm E}^i\,v_{\rm E}^j\,\overline{w}_{,j}({\bf x}_{\rm E})
\biggr)
+{\cal O}(c^{-4}),
\\
\label{t*}
&&T=t^*-{1\over c^2} A(t^*)+{\cal O}(c^{-4}).
\end{eqnarray}
\noindent
For any function $A({\bf x}_{\rm E})$ means $A(t,{\bf x}_{\rm E}(t))$.
Parentheses and brackets around a group indices signify, respectively,
symmetric and antisymmetric parts of the corresponding expressions:
$A^{(ij)}={1\over 2}\,\left(A^{ij}+A^{ji}\right)$ and
$A^{[ij]}={1\over 2}\,\left(A^{ij}-A^{ji}\right)$, etc.  The moment of time
$t^*$ which appears in (\ref{aei}) is defined by (\ref{t*}).  The
function $C(t,{\bf x})$ is not fixed completely, but should only satisfy
(\ref{C:laplace}). It is clear, however, that the PPN equations of
motion do not depend on $C(t,{\bf x})$.  Equations
(\ref{aei})--(\ref{t*}) represent equations of motion of the origin of
the local reference system relative to the global one. For $Q_a=0$ the
equations coincide with the geodesic equations in the external potentials
$\overline{w}$ and $\overline{w}^i$.

The external potentials $W_{\rm T}$ and $W^a_{\rm T}$ represent
the tidal fields of external masses and are given by
\begin{eqnarray}\label{W_T}
W_{\rm T}(T,{\bf X})=&&
{\overline{w}}(t,{\bf x})-{\overline{w}}({\bf x}_{\rm E})
-{\overline{w}}_{,j}({\bf x}_{\rm E})r_{\rm E}^j
\nonumber \\
&&+{1\over c^2}\biggl(
-2(1+\gamma) v_{\rm E}^i
\left({\overline{w}}^i(t,{\bf x})-{\overline{w}}^i({\bf x}_{\rm E})
-{\overline{w}}^i_{,j}({\bf x}_{\rm E})r_{\rm E}^j \right)
+(1+\gamma) v_{\rm E}^2 W_{\rm T}
\nonumber \\
&&
\phantom{+{1\over c^2}\biggl(}
+(1+\gamma)\dot{\overline{w}}^{i,j}({\bf x}_{\rm E})\,r_{\rm E}^i
\,r_{\rm E}^j
+{1\over 2}\,\gamma\,\ddot{{\overline{w}}}({\bf x}_{\rm E})\,r_{\rm E}^2
+\left({1\over 2}-\beta-\gamma\right)\,{(a_{\rm E}^i r_{\rm E}^i)}^2
\nonumber \\
&&
\phantom{+{1\over c^2}\biggl(}
+(1-2\beta-2\gamma)\,Q_aX^a\,a_{\rm E}^ir_{\rm E}^i
-\gamma\,v_{\rm E}^ir_{\rm E}^i\,
{\overline{w}}_{,j}({\bf x}_{\rm E})\,r_{\rm E}^j
+{1\over 2}\gamma\,r_{\rm E}^2\,R^a_{\ i}a_{\rm E}^i Q_a
\nonumber \\
&&
\phantom{+{1\over c^2}\biggl(}
+{\partial\over\partial T} C(T,{\bf X})
+2(1-\beta)\left({\overline{w}}({\bf x}_{\rm E})
+a_{\rm E}^ir_{\rm E}^i\right) W_{\rm T}
\biggr)+{\cal O}(c^{-4}),
\\
\label{U_T^i}
W_{\rm T}^a(T,{\bf X})=&&R^a_{\ i}\,\biggl\{{\overline{w}}^i(t,{\bf x})-
{\overline{w}}^i({\bf x}_{\rm E})
-{\overline{w}}^i_{,j}({\bf x}_{\rm E})r_{\rm E}^j
-v_{\rm E}^i W_{\rm T}(T,{\bf X})
\nonumber \\
&&\phantom{R^a_{\ i}\,\biggl\{}
+{1\over 2(1+\gamma)}\left(
\gamma\,\left(r_{\rm E}^i a_{\rm E}^j r_{\rm E}^j
-{1\over 2}\,a_{\rm E}^i r_{\rm E}^2\right)
-C_{,i}(T,{\bf X})\right)\biggr\}+{\cal O}(c^{-2}).
\end{eqnarray}
\noindent
Finally, the function $\Psi$ reads
\begin{eqnarray}\label{Psi}
\Psi(T,{\bf X})&=&-\eta\, \left(w_{\rm E}(t,{\bf x})
\left({\overline{w}}({\bf x}_{\rm E})+ a_{\rm E}^i r_{\rm E}^i\right)
-\chi^{\rm E}_{\ \,,i}(t,{\bf x})\,a_{\rm E}^i\right)+{\cal O}(c^{-2}),
\end{eqnarray}
\noindent
$\eta=4\beta-\gamma-3$ being the Nordtvedt parameter, and
$\chi^E(t,{\bf x})$ is defined by the integral (\ref{chi}) taken over
the volume $V$ of the central body. The presence of this function
reflects the violation of property {\bf B} in the local reference
system $(T,X^a)$. Property {\bf A} is satisfied, since both $W_{\rm T}$
and $W^a_{\rm T}$ are $\sim {\cal O}({\bf X}^2)$.  One can show that by
changing the transformations (\ref{trans:time})--(\ref{t*}) one can
construct another version of the local reference system where property
{\bf B} is valid, but property {\bf A} is violated. One can also show
that it is impossible to satisfy both properties simultaneously. This
fact is a direct consequence of the violation of the Equivalence
Principle for $\eta\neq0$.

The function $Q_a(T)$ characterizes the world line of the origin of the
local reference system.  $Q_a(T)$ represents the acceleration of the
instantaneous locally inertial reference system (whose origin coincides
with that of the local reference system at a given moment of time)
expressed in the local reference system.  E.g., for $Q_a=0$ the origin
moves along a geodesic in the external gravitational field.  The value
$Q_a$ can be also chosen is such a way that the multipole expansion of
the internal gravitational field $W_{\rm E}$ does not have a dipole
component.  The function $C_a(T)$ describes the spatial orientation of
the local reference system with respect to the global one. Its relation
to the orthogonal matrix $R^a_{\  i}$ is defined by (\ref{Rij-Ca}).
One can choose $C_a(T)$ so that $R^a_{\ i}=\delta^a_{\ i}$ and the
resulting local reference system does not rotate relative to the global
one.  Another possible choice is $C_a=0$ resulting in a dynamically
nonrotating local reference system and the orthogonal matrix $R^a_{\
i}$ in the transformations of the spatial coordinates represents
the well-known  de Sitter, Lense-Thirring and Thomas precessions.

\section{Rotational equations of motions}

The local reference system of an extended massive body described above
allows us to derive the rotational equations of motion of the body in
the framework of the PPN formalism. From the equation
(here, $\cal G$ is the determinant of the local metric tensor)
\begin{equation}\label{pN:fock}
\varepsilon_{abc}\,\int_{V} (-{\cal G})\,X^b\,{\cal T}^{c\beta}_{\ \ ;\beta}
d^3X=0,
\end{equation}
\noindent
which is valid due to the local equations of motion
\begin{equation}\label{local-eq-m}
{\cal T}^{\alpha\beta}_{\ \ \ ;\beta}=0,
\end{equation}
\noindent
in analogy to General Relativity (see \cite{dsx:93,klio:96})
one can derive the rotational equations of motion for the body
\begin{equation}\label{dSdt-ppn}
{d\over dT}\,S^a=L^a_{\rm inertial}+L^a_{\rm T}+L^a_{\rm Nor}
+{\cal O}(c^{-4}),
\end{equation}
\noindent
where the PPN spin $S^a$ is defined by
\begin{eqnarray}\label{pN:spin}
S^a&=&\varepsilon_{abc} \int_{V} X^b\, p^c(T,{\bf X})\, d^3X
+{\cal O}(c^{-4}),
\\ \label{pN:Q}
p^a&=&\Sigma^a\, (1+ {5\gamma-1\over c^2}\, W)
-{1\over 2c^2}\,G\,\Sigma
\int_{V} \Sigma^b(T,{\bf X}')\,\,{(4\gamma+3)\,\delta^{ab} +
n^a n^b \over |{\bf X}-{\bf X}'|}\,\, d^3X'+{\cal O}(c^{-4}),
\\ \label{ni}
n^a&=&{X^a-X'^a\over |{\bf X}-{\bf X}'|},
\end{eqnarray}
\noindent
the PPN inertial and tidal torques read
\begin{eqnarray}
\label{torque:inertial}
&&L_{\rm inertial}^a=\varepsilon_{abc} \int_{V} X^b
\left(
\Sigma\,Q_c+{1\over c^2}\,2(1+\gamma)\,
\varepsilon_{cde}
\left(C_d\,\Sigma^e+{1\over2}\,\Sigma\,\dot C_d\,X^e\right)
\right)
\,d^3X,
\\
\label{torque:tidal}
&&L_{\rm T}^a=\varepsilon_{abc} \int_{V} X^b
\left(\Sigma\,W_{{\rm T},c}+{1\over c^2}\,2(1+\gamma)\,
\left(\Sigma\,{\partial\over \partial T}
W^c_{\rm T}+\Sigma^d\,
\left(W^c_{{\rm T},d}-W^d_{{\rm T},c}\right)\right)\right)
\,d^3X+{\cal O}(c^{-4}),
\end{eqnarray}
\noindent
and the additional torque $L^a_{\rm Nor}$ is defined as
\begin{eqnarray}\label{torque:Nor}
L_{\rm Nor}^a&=&{1\over c^2}\,\eta\,
\varepsilon_{abc} \int_V \Sigma\,X^b\,\Psi_{,c}\,d^3X=
{1\over c^2}\,\eta\,\varepsilon_{abc}\,\Omega_{\rm E}^b\,a_{\rm E}^c
+{\cal O}(c^{-4}),
\\ \label{Omegai}
\Omega_{\rm E}^a&=&-{1\over 2} \int_V \Sigma\  W_{\rm E}\, X^a\, d^3X
+{\cal O}(c^{-2}).
\end{eqnarray}
\noindent
The inertial torque is due to inertial forces dependent on the choice
of $Q_a$ and $C_a$. The PPN tidal torque $L^a_{\rm T}$
is due to external gravitational potentials $W_{\rm T}$ and $W_{\rm
T}^a$.  It vanishes for an isolated body and is proportional to the
square of the spatial size of the body. The additional torque $L^a_{\rm
Nor}$ represents an analogy to the Nordtvedt effect (known from the PPN
translational equations of motion of extended self-gravitating bodies)
in the rotational equations of motion. $L^a_{\rm Nor}$ is proportional
to the Nordtvedt parameter $\eta$, which is not zero in a particular
theory of gravitation only if that theory leads to the violation of the
Equivalence Principle.  The effect results from the difference
between  the center of inertial mass and that of
gravitational mass (the latter involves also the gravitational binding
energy). The Nordtvedt effect consists actually in the fact that the
two forms of mass-energy experience different accelerations provided
that the Strong Equivalence Principle is violated. This fact results in
the additional torque $L^a_{\rm Nor}$. This additional
torque in the rotational equations of motion enables us in principle to
test the Strong Equivalence Principle with observations of rotational
motion of celestial bodies:  if $4\beta-\gamma-3\neq 0$ rotational
motion of a body depends on its acceleration relative to the global
reference system. $L^a_{\rm Nor}$ is proportional also to the integral
$\Omega^a_{\rm E}$ which is an analogy to the internal gravitational
energy of the body $\Omega_{\rm E}$ appearing in the PPN translational
equations of motion. $\Omega^a_{\rm E}$ obviously vanishes for
spherically symmetric bodies. Roughly speaking, for $\Omega^a_{\rm E}$
to be non-zero the body should be non-symmetric with respect to its
center of mass (for example, $\Omega^a_{\rm E}\neq 0$ for a homogeneous
semi-sphere). This implies that $\Omega^a_{\rm E}$ is quite small for
typical celestial bodies.  For that reason it is unclear if this ``new
effect'' appearing in the metric theories of gravity will lead to
measurable consequences.

More details about the local PPN reference system as well as about
various equations of motion and their multipole expansions will be
published elsewhere.

\acknowledgements

S.K. was partially supported by a research fellowship of the Alexander
von Humboldt Foundation. We are grateful to the anonymous
referee for his valuable comments and suggestions.

\end{document}